# First-Class Variability Modeling in Matlab/Simulink

Arne Haber
Software Engineering
RWTH Aachen University, Germany
http://www.se-rwth.de/

Carsten Kolassa
Software Engineering
RWTH Aachen University, Germany
http://www.se-rwth.de/

Peter Manhart
Software-Variantenmanagement
Daimler AG, Germany
http://www.daimler.com

Pedram Mir Seyed Nazari
Software Engineering
RWTH Aachen University, Germany
http://www.se-rwth.de/

Bernhard Rumpe
Software Engineering
RWTH Aachen University, Germany
http://www.se-rwth.de/

Ina Schaefer
Software Engineering and Automotive Informatics
TU Braunschweig, Germany
http://www.tu-bs.de/isf

## ABSTRACT

Modern cars exist in an vast number of variants. Thus, variability has to be dealt with in all phases of the development process, in particular during model-based development of software-intensive functionality using Matlab/Simulink. Currently, variability is often encoded within a functional model leading to so called 150%-models which easily become very complex and do not scale for larger product lines. To counter these problems, we propose a modular variability modeling approach for Matlab/Simulink based on the concept of delta modeling [8, 9, 24]. A functional variant is described by a delta encapsulating a set of modifications. A sequence of deltas can be applied to a core product to derive the desired variant. We present a prototypical implementation, which is integrated into Matlab/Simulink and offers graphical editing of delta models.

## Categories and Subject Descriptors

D.2.6 [**Software Engineering**]: Programming Environments—*Graphical environments*; D.2.2 [**Software Engineering**]: Design Tools and Techniques—*Computer-aided software engineering (CASE)*

## 1. INTRODUCTION

Modern cars are highly configurable. Variability extends over the whole range of vehicle functionality from technical base functionality to comfort functionality in the interior. Explicit management of variability is essential because many market-specific regulations change and lead to agile adaptions for requirements of different car lines. Also, the large number of variants is considered one of the main success factors of car manufacturers, as it allows customers to tailor their car to their own requirements and preferences, e.g. comfort, safety, or sportiness. However, in order to realize this variability the different variants of the vehicle functions have to be planned and realized during development. This requires to be able to deal with variability in all development phases by adequate means.

In this paper, we concentrate on managing variability for the model-based development of variant-rich vehicle functions in Matlab/Simulink [32]. Simulink allows modeling a dynamic system using block diagrams. Blocks may be hierarchical decomposed using subsystem blocks to model layered architectures of vehicle functions and may be connected to model inter-block communication. Behavior is either created by composing atomic blocks that, e.g., realize mathematical functions, by composing hierarchical blocks, or by modeling state charts in Stateflow [33].

While variability modeling on the requirements level with feature models [17] is well understood, the representation of different functional variants in the model-based development using Matlab/Simulink is still problematic. The reason is that there are no first-class variability modeling concepts in Matlab/Simulink so far. Instead, an annotative variability modeling approach [30] is used where a model contains all Simulink blocks that may be contained in any variant such that it is also called 150%-model. The selection of blocks for different variants is realized by an encoding with model elements that are actually to be used for modeling functionality. For instance, switch-blocks or if-action-blocks which are intended to capture the selection of functionality at runtime are used for the modeling of variants. The conditions of the switch- or if-action block are identified with a product feature and can be set externally by assigning a value to a constant which corresponds to the selection of a variant. Fig. 1 depicts an example for variability encoding using if-action blocks based on a braking system. Variable block `BASE` is set externally and is forwarded to an if block. If `BASE` is selected (u1 == 1), then if-action block `PressureCalculator` is selected that provides basic functionality by calculating the brake pressure for each wheel based on a brake signal. If `BASE` is not selected, if-action block `ABS` is used that also takes the speed of each wheel as input to additionally prevent wheels from blocking during a slowing-down process. Outgoing ports are restricted to receive signals from one receiver exclusively, hence the outputs of `PressureCalculator` and `ABS` have to be merged by `Merge` blocks.

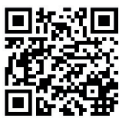

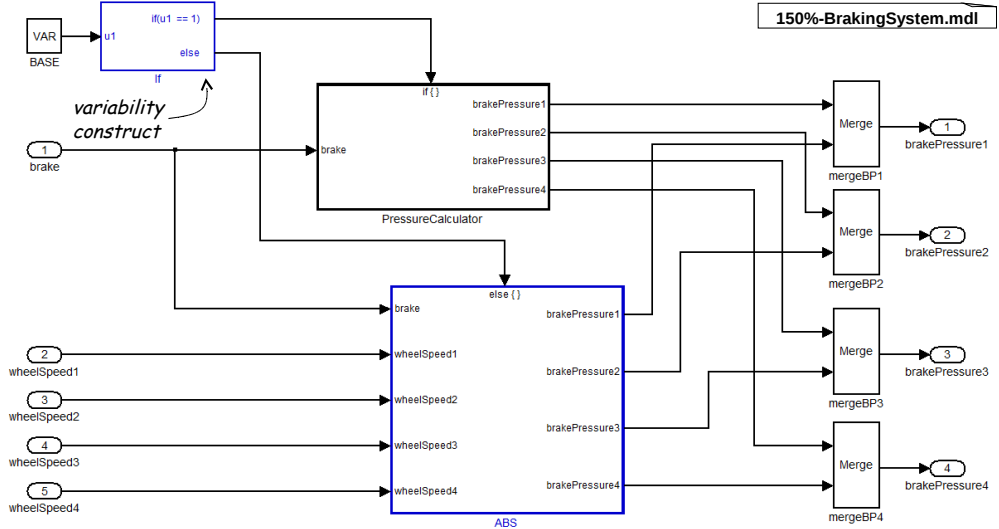

Figure 1: Variability Encoding in Current Practice

This approach to encode variability by functional blocks leads to a significant increase in model complexity, since different functional variants are encoded in the same 150%-model. Additionally, descriptions of functionality and variability are contained in the same model violating the principle of separation of concerns. In these models, the developer can no longer concentrate on designing the functionality, but the modeling is bound to the representation of functionality. For complex functionality with many variants, the resulting model easily becomes very large and difficult to debug. Particular problems arise when variants require changes in several places of a model or on different hierarchical layers. This leads to a variability encoding that is distributed all over the model such that requirements of specific variants can barely be traced back to the variable parts of the model.

To counter these problems, our goal is to extend Matlab/Simulink with an explicit first-class variability modeling concept that support modular and hierarchical variability modeling for dealing with large and complex variant-rich systems. In particular, we aim at an integration of this approach into the existing development processes and tool infrastructure within Matlab/Simulink. To achieve this goal, we extend Matlab/Simulink by the concept of delta modeling [3]. Delta modeling is a flexible transformational approach for modular modeling of variability. It has been successfully applied to class diagrams [23], architecture description languages [9, 8], or programing languages [24, 26]. The main idea of delta modeling is to describe a set of variable systems by a designated *core model* and a set of *deltas* that specify modifications to the core system to obtain other product variants. A particular system variant is obtained automatically by applying a set of deltas in a specific order to the core system. The delta modeling approach allows encapsulating variability within deltas that are first-class variability modeling elements.

In this paper, we present a delta modeling extension for Simulink, which we call *Delta Simulink*. Our contributions are the following: Delta Simulink is the first delta modeling approach using a graphical interface while previous delta modeling approach use textual languages. It is fully integrated in Simulink allowing to modify, create and apply deltas directly from the graphical user interface. It is implemented based on the tool chain for delta modeling for block-oriented functional architectures in ∆-MontiArc [9, 8]. The applicability is demonstrated by the example of a product line of braking systems.

The paper is structured as follows: In Sect. 2, we describe the concepts Delta Simulink, and in Sect. 3, its prototypical realization. Sect. 4 illustrates Delta Simulink with an example. Sect. 5 discusses the applicability of Delta Simulink in industrial practice. Sect. 6 reviews related work and Sect. 7 concludes this paper.

## 2. DELTA SIMULINK

In order to obtain a delta modeling extension in Simulink, called *Delta Simulink*, we have to define the modeling language for the core model, the possible set of delta operations, the language for the application order constraints, the variant selection and the variant generation process.

**Core Models.** Core models are defined using plain Simulink models. Our prototype of Delta Simulink restricts the set of supported modeling elements to `subsystem` blocks, `model` blocks, `connections`, `inport`, and `outport` blocks. Subsystem blocks hierarchically decompose a system. Model blocks include a referenced model into the current model. The interface of all block types is given by in- and outports. Signal flows are modeled using connections between ports. An example for a valid core model is give in Fig. 2. It depicts a model of a braking system component. The component receives a brake command on its port `brake`. This signal is forwarded to the model block `brakefunction` that references the model `PressureCalculator`. The contained block calculates the brake pressure for each wheel and emits this results via its ports `brakePressure1...brakePressure4`. These results are forwarded to the outgoing ports of the braking system. The model corresponds to the basic braking system variant contained in Fig. 1.

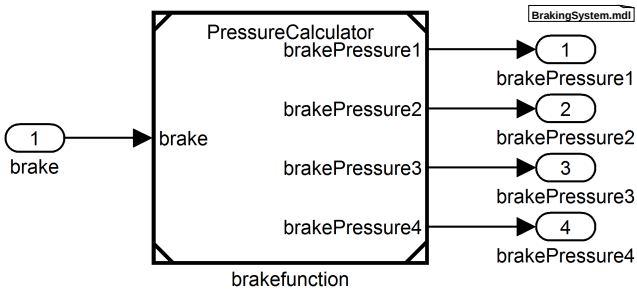

Figure 2: Simulink Model of Core Braking System

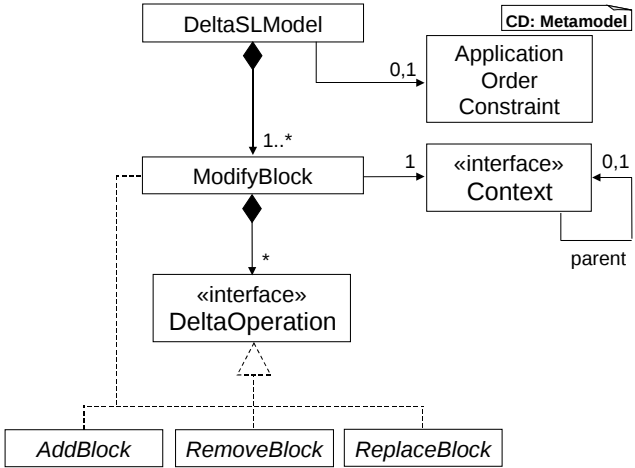

Figure 3: Delta Simulink Meta Model

| Operation | Affected elements |
|---|---|
| add | model references, subsystems, ports, connections |
| remove | model references, subsystems, ports, connections |
| modify | models, subsystems |
| replace | model references, subsystems |

Table 1: Overview of Delta Operations

**Delta Operations.** To specify deltas, language constructs are needed to define delta operations on the aforementioned model elements. The syntactical structure of deltas in Delta Simulink is given by the meta-model depicted in Fig. 3. Deltas are defined in their own Simulink model file (DeltaSLModel) that contains at least one ModifyBlock at the top layer that references the context that should be modified. A Context corresponds to either a model or a subsystem block and may be hierarchically decomposed itself. This way subsystems contained in a model can be modified. In addition, a delta may be associated with an Application Order Constraint ($AOC$) that is used to explicitly model inter-delta dependencies. In the majority of cases this is needed if deltas modify the same (core) model elements or product features realized by deltas depend on each other. For example, a delta for adding a brake function without ABS and a delta for adding a brake function with ABS are mutually exclusive (if it should not be possible to switch ABS on and off). So both deltas need the constraint that they cannot be applied after the other one. Modifications of the context are specified by DeltaOperations including Add-, Remove-, Replace-, and ModifyBlocks.

Table 1 contains an overview of the delta operations that are available in Delta Simulink and the model elements on which they can be applied. All model elements supported by our prototype may be added or removed from the enclosing block. Additional modification operations are available for hierarchically decomposed elements, i.e., model reference blocks and subsystem blocks. A modify operator allows altering the internal structure of these blocks by defining a set of delta operations for changing the contained block elements. We also introduce a replace operator for hierarchically decomposed elements. This operator substitutes a block $bl$ with another block $newBl$ that has a compatible interface. The interface of $newBL$ is compatible if $newBl$ has at least the same incoming and outgoing ports. By the replace operator, all connections from/to $bl$ are removed, $bl$ itself is removed, a block $newBl$ is created, and in addition all previously existing connections are created such that now $newBl$ is connected instead of $bl$.

**Concrete Delta Syntax.** Delta operations are specified using a custom Simulink context menu. If a model at the top level should be modified, the name of the respective model block has to start with modify model followed by the model's name. If a subsystem should be modified, the name of the respective subsystem block has to start with modify followed by the subsystem's name. Modify blocks hierarchically contain DeltaOperations that transform elements of the associated context. As shown in the meta model (cf. Fig. 3), the delta operations can be Add-, Remove-, Replace-, and ModifyBlocks. A block or line that is added is a delta is highlighted green. If it is removed, it is highlighted in red.

A ReplaceBlock is represented by a subsystem or model reference block with an orange outline, depending on the model element that should substitute the target block. The name of a replace block consists of several parts. After the keyword replace the element to be replaced is referenced by its name. Then, the keyword with is followed by the substitute, that may be either a single block name if the substitute is a subsystem, or a model name and a block name if the substitute is a model.

An example of a delta in Delta Simulink that adds anti-lock braking functionality to the braking system is depicted in Fig. 4. The top level of the delta DABS is given on the left side. The depicted modify block defines the context of the contained delta operations. Hence, all contained operations affect the model BrakingSystem. The associated AOC states that delta DTW_post must not be applied before the modeled delta, because otherwise changes performed by DTW_post (cf. Fig. 9) would be reverted. In contrast to the meta-model, the AOC is bound to the modify block instead of to the delta itself. This is due to technical limitations of Simulink which does not allow to attach a constraint to the model itself. As a delta may contain more than one modify model block on the top level, it is possible to add more then one AOC to a delta. In this case, all AOCs are combined to a single AOC using a logical AND operation. The delta operations in the right part adds the ports `wheelSpeed1 ... wheelSpeed4` to model `BrakingSystem` and replaces the contained block `brakefunction` with a model reference block that references model `ABS` and has the same block name.

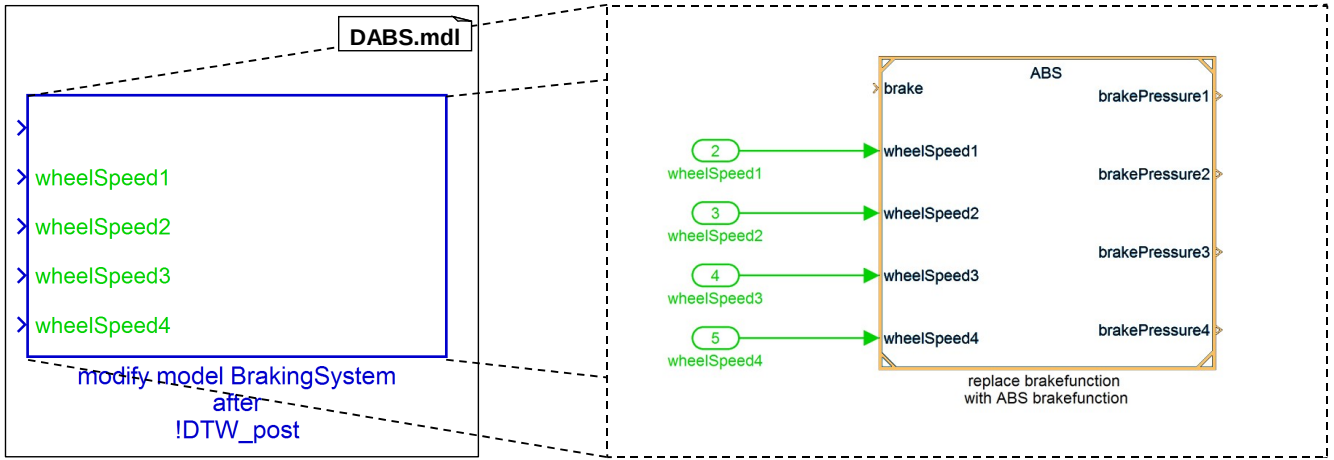

Figure 4: Delta for ABS Functionality

**Application Conditions.** Delta operations must fulfill a number of application conditions to be applicable in a product generation process:

1. If an element $e_{add}$, i.e., a block, ports or a connection, is to be added, there must not exist another element $e$ with the same name in the current context.

2. Connectors may only be added, if (a) source and target exist, and (b) the target is not already a target of another connection.

3. Ports of subsystem blocks may only be removed, if they are unconnected.

4. If an element $e_{rem}$ is to be removed, $e_{rem}$ has to exist in the current context. As a special case, a weak remove is not rejected, if $e_{rem}$ does not exist. A weak remove is useful to ensure that element $e_{rem}$ does not exist after applying a delta.

5. For modify blocks, we require the following:

   (a) The context $c$ of a modify block at the top level of a delta always has to be a model.
   (b) The context $c$ of a modify block at lower levels of a delta always has to be a subsystem.
   (c) The context $c$ of a modify block has to be valid. If $c$ refers a model, this model has to exist. If $c$ refers a subsystem $sub$, $sub$ has to exist in the parent context $c.parent$ of $c$ (c.f. Fig. 3).

6. For an operation "replace $bl$ with $(modName)$ $newBl$", we require:

   (a) The block $bl$ that is to be replaced must exist in the current context.

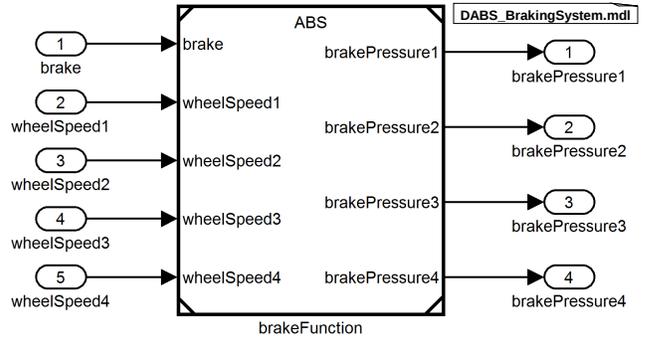

Figure 5: Braking System with ABS

   (b) There must not exist another element named $newBl$ in the current context after removing $bl$.
   (c) A model named $modName$ must exist, if the substitute is a model block.
   (d) The port interface of $bl$ and $newBl$ have to be compatible.

**Variant Selection and Generation.** A product variant is defined by a set of deltas that have to be applied to the core to generate this product variant. The generation process takes this set of selected deltas, called product configuration, as input and computes the order in which the selected deltas have to be applied by interpreting the application order constraints. Then the deltas are applied stepwise in the computed order to the core model. When a delta is applied, the contained deltas operations transform the core or the intermediate model. As an example, the product variant "BrakingSystem with ABS" is defined by the product configuration $\{DABS\}$, containing delta `DABS` as single element. The generated product variant that is created by applying the delta `DABS` to the core model is depicted in Fig. 5.

## 3. PROTOTYPICAL REALIZATION

In this section, we describe the prototypical realization of Delta Simulink as an extension to Matlab/Simulink and explain how a new Simulink variant model is created. The prototypical implementation of Delta Simulink is based on the existing language implementation Δ-MontiArc which we have developed in previous work [9, 8, 10]. Δ-MontiArc implements the concept of delta modeling for the architecture description language MontiArc [12] that uses concepts similar to Simulink block diagrams. However, Δ-MontiArc is a purely textual, prototypical language, whereas Delta Simulink is the first graphical delta modeling language and integrated into the industrial-scale development environment Matlab/Simulink.

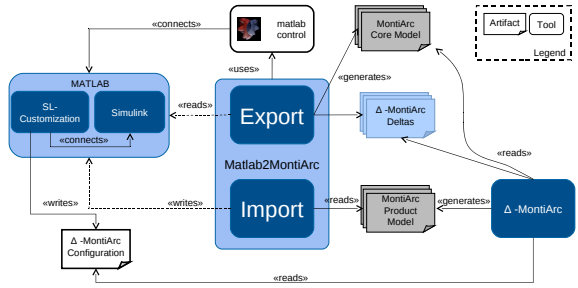

**Figure 6: Architecture of Delta Simulink Prototype**

Fig. 6 shows the architecture of the prototypical implementation of Delta Simulink. Simulink core models and delta models are both created graphically within Simulink following the concepts described in Sect. 2. Core models are defined as standard Simulink models. Deltas are specified within an own delta modeling mode in Simulink, and each delta is stored in a separate file.

The switching between the two modes is done via the Tools menu. In the normal mode colors can be used without limitations, while in the delta modeling mode elements are highlighted as shown in Fig. 7. More precisely, blocks that should be modified are marked as *modify* and are highlighted in blue. Blocks marked for an *add* operation are highlighted in green, while blocks highlighted in red correspond to *remove* operations. Blocks highlighted in orange represent *replace* operations. The delta modeling mode is implemented using the *sl_customize* API of Simulink.

Using the Matlab Control Library [31], Simulink model construction commands can be sent to Matlab in order to read Simulink core models and the delta models. In this way, Simulink core model elements are transformed to their corresponding MontiArc elements and Simulink deltas are translated to Δ-MontiArc deltas.

To specify which deltas are required to generate a particular product variant, a configuration file with the set of required deltas has to be provided. The configuration file can be created automatically via a tool menu in Simulink. With the (transformed) MontiArc core models, the Δ-MontiArc deltas, each stored in separate file, and the Δ-MontiArc configuration file as input, Δ-MontiArc determines the application order and handles conflicts and model dependencies automatically. It does this by using application order constraints (*AOC*) that define which deltas have to be applied before or must not be applied before the respective delta. In addition Δ-MontiArc also checks correctness of generated models according to the context conditions described in [12]. Finally, the generated variant is transformed back to a Simulink model by an import component using the Matlab Control Library.

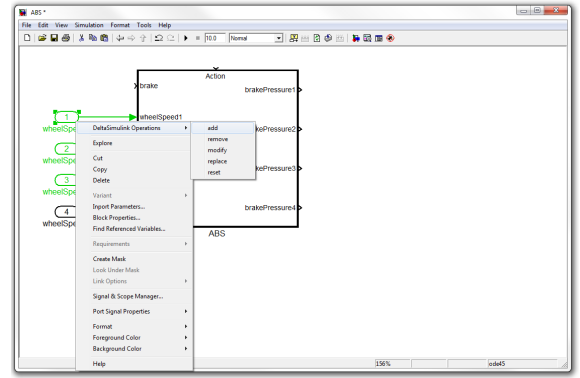

**Figure 7: Delta modeling mode in Simulink.**

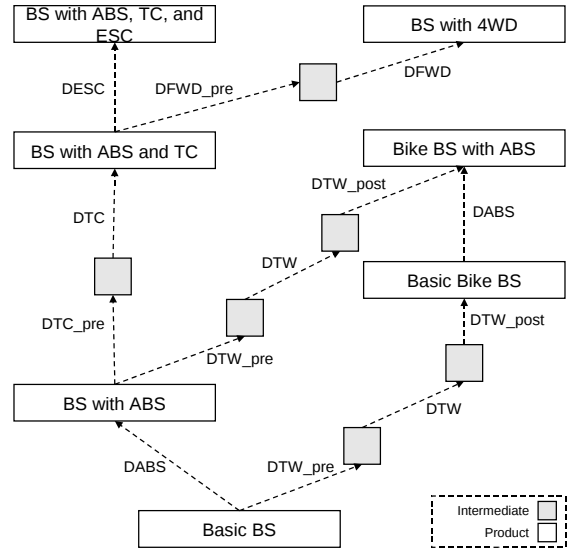

**Figure 8: Overview of the Braking System PL**

## 4. CASE EXAMPLE

To evaluate the applicability of Delta Simulink, we realized a braking system product line. An overview of the product variants and the corresponding set of deltas is given in Fig. 8. The product line consists of seven product variants. Starting from a basic braking system that corresponds to the core model (see Fig. 2), further product variants are derived by adding and combining additional functionality like antilocking (`ABS`), traction control (`TC`), electronic stability control (`ESC`), or four wheel drive (`FWD`). In addition, variants for motorbikes (`TW`) with and without ABS are included. Fig. 8 also contains the delta structure tree with the deltas and their application order to generate the different product variants. If, for example, product "BS with ABS and TC" should be generated, the deltas `DABS`, `DTC_pre`, and `DTC` have to be applied to the core model.

The number of deltas is not necessarily equal to the number of features, since additional deltas may have to be applied before or after the delta that realizes the actual feature. In our case study, we denote these deltas by a suffix `_pre` or `_post`. Usually, these deltas are necessary if more than one delta operation affects a specific model element.

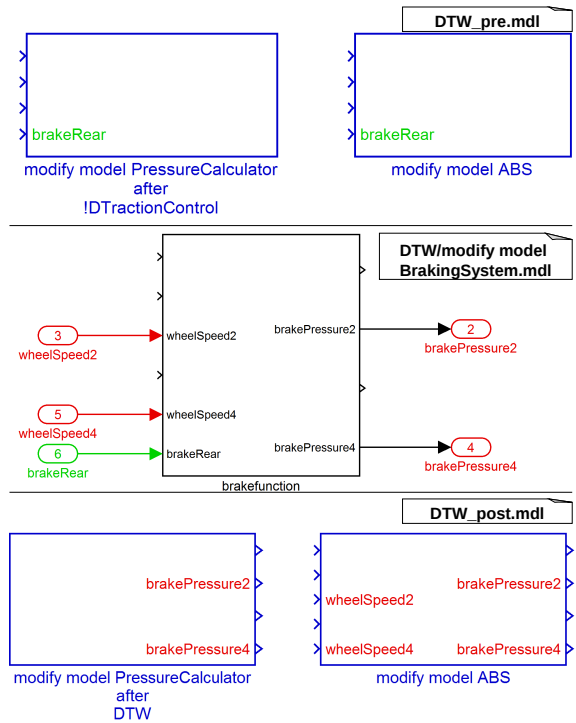

**Figure 9: Deltas for Two Wheel Braking Systems**

In a graphical modeling language, such as Simulink, an ordering of these modification operations cannot be specified using the graphical formalism, but the ordering is necessary to generate a valid product. In this case, the modification operations are split into several deltas that are explicitly ordered by their application order constraints. In addition, some restrictions in Simulink itself prevent capturing delta operations that affect the same model element in a single delta. For example, connections have to be removed before a model block can be rewired, because a port can exclusively be a target of one connection. So, removing a connection to a port and adding a new connection to the same port in one single delta is not feasible. In this case, the removal operation and the addition operation are represented in separate deltas that are successively applied.

An example for such a sequence of deltas is given in Fig. 9. The three deltas `DTW_pre`, `DTW`, and `DTW_post` generate the braking system for motor bikes. These deltas can be applied independent of the application of the delta for the ABS feature. The first delta `DTW_pre` adds an incoming port `brakeRear` to the Simulink models `PressureCalculator` and `ABS`. The content of the modify model block defined in delta `DTW` that transforms model `BrakingSystem` is depicted in the middle of Fig 9. It removes the incoming ports `wheelSpeed2` and `wheelSpeed4`, the outgoing ports `brakePressure2` and `brakePressure4` as well as the connections to and from block `brakefunction`. Here, a *weak remove* is used, because `wheelSpeed2` and `wheelSpeed4` only exist, if delta `DABS` has been applied before (see Fig 4). In addition, an incoming port `brakeRear` and its connection to port `brakeRear` of block `brakefunction` is added. It is necessary to split up `DTW_pre` and `DTW`, because according to the application conditions given in Sect. 2 the target port `brakeRear` of the connection added in `DTW` has to exist. Finally, delta `DTW_post` removes the unused ports of the `PressureCalculator` and `ABS` models, since according to the application conditions these ports may only be removed after the connections to and from them have been removed. The result of configuration {DTW_pre, DTW, DTW_post} that defines product "Basic Bike BS" is shown in Fig. 10. It now calculates brake signals based on front and rear brake signals for two wheels only.

## 5. DISCUSSION

Delta Simulink allows representing variability as first-class modeling elements by delta operations which are completely integrated within the Simulink modeling environment. Variability is encapsulated in delta models while functionality is developed in standard Simulink models supporting the separation of concerns principle. The delta operations are completely integrated within the hierarchy of Simulink models such that hierarchical modeling is supported. As Delta Simulink is based on the standard Simulink language, model reuse techniques such as model references are supported as well. Deltas are modeled modularly with explicitly defined dependencies to other delta models. This way distributed development of distinct product variants is supported. Delta modeling inherently supports the automated generation of specific product variants requiring no manual intervention.

The prototypical implementation demonstrates that Delta Simulink can be integrated into the existing Simulink development tool chain. It can be combined with currently used configuration tools which can be used to derive the delta selection for a specific product variant automatically. As Delta Simulink product variants are standard Simulink models, code can be generated using the standard code generators.

However, the prototypical realization needs to be improved in order to facilitate the usage of Delta Simulink in industrial practice. Because of the graphical modeling and the restrictions of the Simulink editor (e.g., not allowing two ingoing connections for the same port which would be inconsistent), some modification operations have to be split up into several deltas to be applied in a sequence. If the Simulink editor can be extended to allow also inconsistent Simulink models in deltas, this problem can be alleviated. In the current version of the prototype, it is assumed that the deltas are specified directly in the Simulink editor by highlighting model elements. In order to simplify the specification of deltas in industrial practice, we consider two possibilities which can be implemented in future versions of Delta Simulink: first, to derive a delta from the differences between two Simulink models; and second, to record a delta from the modifications applied to a Simulink model within the editor.

## 6. RELATED WORK

For modeling variability of block-oriented architectures, there basically exist four different variability modeling approaches (which can be combined partially): annotative, compositional, hierarchical and transformational variability modeling.

*Annotative* approaches represent all product variants in one 150%-model. By removing parts of the model, concrete product models can be derived. Variant annotations define these parts with the help of, e.g. UML stereotypes [35, 7] or presence conditions [4]. Orthogonal variability models (OVM) [22] separate the representation of the model vari-ability and the artifact model. This idea is specialized for architectural models in the variability modeling language (VML) [19]. Opposed to our approach, annotative variability modeling provides no first-class variability modeling concepts such that it hardly scales for large and complex variant-rich systems. Because variability and functionality are mixed in the same model, reuse of common functionality often is not possible. Variants are created using clone-and-own practices.

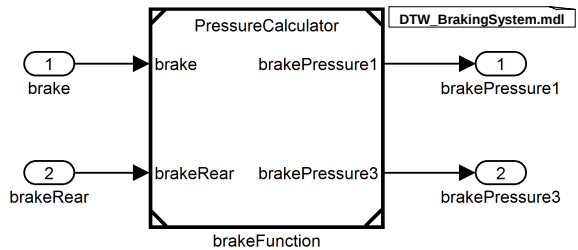

Figure 10: Basic Bike Braking System

*Compositional* approaches associate model fragments with product features that are composed for a particular feature configuration. In [5], product variants are defined by merging model fragments. In [14, 30, 20], aspect-oriented composition is used to build the models whereas feature-oriented model-driven development (FOMDD) [27] combines feature-oriented programming (FOP) with model-driven engineering. Model superposition [1] is another way for composing model fragments. Different from delta modeling, compositional variability modeling is restricted to the addition of elements to a model such that it is always required to start with the smallest possible core models. In contrast, deltas can also remove model elements which allows to start from any existing core model.

*Hierarchical* variability modeling combines the component hierarchy in the architecture with the component variability. In [21], partially defined components are extended with variation points and associated variants where variants can be cross- or non-cross-cutting architectural concerns that are composed with the common component architecture by weaving mechanisms. The extended components are called plastic partial components. In the Koala component model [28, 29], switch components serve as variation points which allow the selection of different subcomponent variants. In [11], the architectural description language MontiArc [12] is extended by hierarchical variability modeling concepts similar to the Koala approach targeting at the architectural design phase, where as Koala is mainly tailored for the implementation phase. Hierarchical variability modeling concepts require appropriate tool support to deal with their inherent complexity.

*Transformational* approaches employ model transformations for capturing product variability. The common variability language (CVL) [13] represents the variability of a base model by rules describing how modeling elements of the base model have to be substituted in order to obtain a particular product model. In [16], graph transformation rules capture the variability of a single kernel model comprising all commonality. For describing variability in software archtictecture often a combination of the above approaches are used. In [15], architectural variability is represented by change sets containing additions and removals of components and component connections that are applied to a base line architecture.

The concept of delta modeling [3, 24, 10] that is applied to represent variability in Delta Simulink can also be classified as a combination of a compositional, hierarchical and transformational approach. The deltas are defined in separate models and can transform also blocks on lower hierarchical levels. The core model is transformed to a new variant by applying a set of deltas. Delta modeling allows modular, yet flexible variability modeling in an intuitive way, such that we decided to base Delta Simulink on it.

With respect to Matlab/Simulink, we have so far only observed variability modeling approaches using 150%-models. [34] presents a decision-oriented approach for modeling variability in a prototypical Matlab clone. Common functionality has to be modeled first and is then extended with explicit variation points within the same model. A variant is created by answering predefined questions which resolve the variation points to given variants. In contrast, in Delta Simulink, the specification of variation points in the core model is not needed due to the use of deltas. In [2, 6], variation points containing variability information for Simulink models and variability mechanisms determining how variability is resolved are distinguished. For the latter, resolution blocks are used, e.g., model variant and enabled subsystem blocks. Depending on the input signal of a resolution block, a specific variant is chosen. The input signal is regulated by so called control blocks, i.e., constant blocks, referencing the variant parameter and the value zero (variant not selected) or one (variant selected). The variant blocks have a reference to the corresponding variation point. Hence, for a specific variation point, several variant mechanisms can exist which can be accessed through a generic interface. In contrast to our approach, in [2, 6] variability and functional aspects are represented in the same model.

## 7. CONCLUSION AND FUTURE WORK

In this paper, we apply the concept of delta modeling [3] to Matlab/Simulink in order to obtain a modular, yet flexible variability modeling concept for Simulink models. Delta Simulink provides first-class language constructs to represent variability. This provides a clear separation between modeling functionality and variability and alleviates the complexity of modeling complex variant-rich functionalities with Matlab/Simulink. It is the first graphical delta modeling language and integrated into the industrial-scale model-based development environment Simulink. For future work, we aim at extending the language constructs of Delta Simulink by computation blocks and by fine-grained delta operations for busses. Along the lines of [18], we will also extend delta modeling to Stateflow models within Simulink block diagrams. We plan to develope a consistency checker (following [25]) and a tool for managing and debugging deltas. We also intent to evaluate our prototype in industrial practice.